\documentclass[12pt, reqno]{amsart}
\usepackage{hyperref}

\begin{document}
\title[\hfilneg \hfil Green's Function for LFD with Variable Coefficients]
{Explicit Representation of Green's Function for Linear Fractional Differential Operator with Variable Coefficients}

\author[Myong-Ha Kim, Hyong-Chol O\hfil
\hfilneg]
{Myong-Ha Kim, Hyong-Chol O}

\address{Myong-Ha Kim \newline
Faculty of Mathematics, {\bf Kim Il Sung} University , Pyongyang, D. P. R. Korea}
\email{myongha\_kim@yahoo.com}

\address{Hyong-Chol O  \newline
Faculty of Mathematics, {\bf Kim Il Sung} University , Pyongyang, D. P. R. Korea}
\email{ohyongchol@yahoo.com}

\thanks{First submitted Aug 9, 2012  last version revised Sep 3, 2013. Accepted to Journal of Fractional Calculus and Applications, 5(1) 2014.}

\setcounter{page}{1 }

\maketitle

\begin{abstract}
We provide explicit representations of Green's functions for general linear fractional differential operators with {\it variable coefficients} and Riemann-Liouvilles derivatives. We assume that all their coefficients are continuous in $[0, \infty)$. Using the explicit representations for Green's function, we obtain explicit representations for solution of inhomogeneous fractional differential equation with variable coefficients of general type. Therefore the method of Green's function, which was developed in previous research for solution of fractional differential equation with constant coefficients, is extended to the case of fractional differential equations with {\it variable coefficients}.
\end{abstract}

\section{Introduction}
It seems that the concept of fractional Green’s functions for fractional differential operators have been introduced by S.I.Meshkov \cite{mes} for the first time in 1974 to represent the solutions of inhomogeneous fractional differential equation with constant coefficients and single term. This concept is one that is extended from the concept of Green's function for ordinary differential operator with natural number order given by M.A.Naimark \cite{nai} in1969 to fractional (real number) order.\\\indent
After fractional Green's function have been studied by S.I. Meshkov in 1974, many authors have derived explicit representation for Green's functions of linear fractional differential operators with constant coefficients \cite{HLT, HLL, MS1, MS2, pod1, pod2}. With the help of Green's function and some special functions such as Mittag-Leffler function, in 1993 Miller and Ross in \cite{MR} obtained the explicit representations of solutions of some classes of homogeneous linear fractional differential equations FDEs. In 1994, I. Podlubny derived an explicit representation for Green's function of an arbitrary linear fractional differential operator with constant coefficients by using Laplace transform in \cite{pod1}. Hu Y. et al. \cite{HLL} in 2008 provided a representation formula of Green's function for the above mentioned fractional differential operators with constant coefficients by Adomian decomposition method to apply to representations of the non-homogeneous fractional differential equations. Morita and Sato in \cite{MS1} gave a representation formula of Green's functions for initial value problem of fractional differential operators with constant coefficients by the Neumann series. Bonilla and Junshong \cite{BRT} provide an explicit representation for solution of system fractional differential equations with constant matrix coefficients and single term. X. Huang et al. \cite{HL} provided an explicit representation of Green's function for fractional differential operator with constant coefficients. 

A. A. Kilbas et al. \cite{KRRT} presented a method of solving fractional differential equations with variable coefficients in the neighborhood of ordinary point by power series method.

From the summarizing above we can say that several authors provided explicit representation formula of Green's function for fractional differential operators with constant coefficients but we couldn’t find out the results on arbitrary linear fractional differential operators with variable coefficients.

In this paper we derived an explicit representation formula of Green's function for arbitrary linear fractional differential operators with continuous coefficients and Riemann-Liouville fractional derivatives and applied it to get solution representation of inhomogeneous fractional differential equation. Therefore the method of Green's function which was developed for solution of fractional differential equation with constant coefficients in previous research is extended to the case of fractional differential equations with variable coefficients.

\section{Definitions and Preliminaries}
{\bf Definition 2.1} \cite{KST} For a real number $\gamma(0\leq\gamma\leq 1)$ and $n\in N$,  we define as follows:
\[C_\gamma^n[a,\infty):=\{f:[a, \infty) \rightarrow \mathbb{R}:(t-a)^\gamma f^{(n)}(t)\in C[a,\infty)\},\]
\[C_\gamma[a,\infty):=C_\gamma^0[a,~\infty).\]

{\bf Definition 2.2} \cite{KST} Let $\alpha > 0, f\in C_\gamma[a,\infty)$. The {\it Riemann-Liouville left-side fractional integral} $I^{\alpha}_{a+}f$ of order $\alpha$ with {\it original at the point} $a$ is defined by
\begin{equation}
I^\alpha_{a+}f=\frac{1}{\Gamma(\alpha)}\int_a^t\frac{f(\tau)}{(t-\tau)^{1-\alpha}}d\tau~,~t>a, \label{1}
\end{equation}
provided the integral exists. Here  $\Gamma(\alpha)$ is the Gamma function, and $I^\alpha_{a+}$ is called an {\it integral operator of order} $\alpha$.

{\bf Definition 2.3} \cite{KST} Let $n-1\le\alpha\le n, n\in \mathbb{N}$ and $I^{n-\alpha}_{a+}f\in C^n_\gamma[a, \infty)$. The {\it Rimann-Liouville fractional derivative} $D^\alpha_{a+}f$ of order $\alpha$ with {\it original at the point} $a$ is defined by
\begin{equation}
D^\alpha_{a+}f(t)=D^nI^{n-\alpha}_{a+}f(t)=\left(\frac{d}{dt}\right)^n\frac{1}{\Gamma(n-\alpha)}\int_a^t\frac{f(\tau)}{(t-\tau)^{\alpha-n+1}}d\tau~,~t>a  \label{2}
\end{equation}
and $D^\alpha_{a+}$ is called the {\it fractional differential operator of order} $\alpha$.

{\bf Definition 2.4} \cite{KST} For $n\in \mathbb{N}$, we denote by $AC^n[a, b]$ the space of complex-valued function $f(x)$ which have continuous derivatives up to order $n-1$ on $[a, b]$ and $f^{(n-1)}(x)$ is absolutely continuous, i.e. the function $f(x)$ for which there exists (almost everywhere) a function $g(x)\in L_1[a,b]$ such that
\[f^{(n-1)}(x)=f^{(n-1)}(0)+\int_a^xg(t)dt.\]
In this case we call $g(x)$ the (generalized) $n$-th derivative of $f(x)$  on $[a, b]$ and simply write $g=f^{(n)}$. In particular we denote $AC^1[a, b]=AC[a, b]$. Then  we can write as follows:
\begin{equation}
AC^n[a, b]=\{f:[a, b]\to \mathbb{C}:~D^{n-1}f(t)\in AC[a, b],~D=\frac{d}{dt}\} \label{3}
\end{equation}
 Here $\mathbb{C}$ is the set of complex numbers.

%
{\bf Lemma 2.1} \cite{KST} {\it The space $AC^n[a, b]$ consists of those and only those function $f(t)$ which be represented in the form}
\begin{equation}
f(t)=(I^n_{a+}\varphi )(t)+\sum_{k=0}^{n-1}C_k(t-a)^k~,\label{4}
\end{equation}
{\it where $\varphi \in L(a, b),~C_k(k=0, 1,\cdots, n-1)$ are arbitrary constants and}
\begin{equation*}
(I^n_{a+}\varphi)(t)=\frac{1}{(n-1)!}\int_a^t(t-\tau)^{n-1}\varphi(\tau)d\tau .
\end{equation*}

{\bf Lemma 2.2} \cite{KST} {\it Let $n\in \mathbb{N} _0={0, 1, \cdots}$ and $\gamma \in \mathbb{R} (0\le\gamma \le 1)$. The space $C^n_\gamma[a, b]$ consists of those and only those functions $f$ which are represented in the form}
\begin{equation}
f(t)=\frac{1}{(n-1)!}\int_a^t(t-\tau )^{n-1}\varphi (\tau )d\tau +\sum_{k=1}^{n-1}C_k(t-a)^k~, \label{5}
\end{equation}
{\it where $\varphi \in C_\gamma (a, b)$ and $C_k(k=0, 1,\cdots, n-1)$ are arbitrary constants.}

{\bf Definition 2.5} \cite{SKM}[14] Let $\alpha >0 ~,~1\le p \le \infty$. The space of functions $I^\alpha_{a+}(L_p)$ are defined by
\begin{equation}
I^\alpha_{a+}(L_p):=\{f:f=I^\alpha_{a+}\varphi,~\varphi\in L_p(a, b)\},\quad I^\alpha_{a+}(L):=I^\alpha_{a+}(L_1).\label{6}
\end{equation}

{\bf Lemma 2.3}  \cite{KST} {\it Let $\alpha>0,~n=[\alpha]+1$ and $f_{n-\alpha}(t):=(I^{n-\alpha}_{a+}f)(t)$ be the fractional integral of order $n-\alpha$ of $f$.}

(a)~{\it If $1\le p \le \infty$ and $f\in I^\alpha_{a+}(L_p)$, then the following equality holds:}
\begin{equation}
(I^\alpha_{a+}D^\alpha_{a+}f)=f(t).\label{7}
\end{equation}
\quad (b)~{\it If $f\in L_1(a, b)$ and $f_{n-\alpha} \in AC^n[a, b]$ then the the following equality holds almost everywhere on $[a, b].$}
\begin{equation}
I^\alpha_{a+}D^\alpha_{a+}f(t)=f(t)-\sum_{j=1}^{n}\frac{f^{(n-j)}_{n-\alpha}(a)}{\Gamma(\alpha-j+1)}(t-a)^{\alpha-j}. \label{8}
\end{equation}

For more detail statements of concepts and properties of fractional calculus, see \cite{KST, MR, pod1, SKM}.

\section{Analytic Representation of Green's Function}
Let us consider the initial value problem (IVP) for fractional differential equations(FDE) given by
\begin{equation}
L(D_{0+})y(t)=h(t)~,~t>0,\label{9}
\end{equation}
\begin{equation}
D^{\alpha_0-j}y(0)=0,~j=1, 2, \cdots, n_0. \label{10}
\end{equation}
Here $L(D_{0+}):=D^{\alpha_0}_{0+}+\sum_{h=1}^{m}a_h(t)D_{0+}^{\alpha_k}$;~$
\alpha_0>\alpha_1>\cdots>\alpha_m \ge 0$~;~$a_h\in C[0, \infty)$ and $D^{\alpha_h}_{0+},~h=0, 1, \cdots, m$ is the Riemann-Liouville left-sided fractional differential operator with the original at $t=0;~ n_0-1<\alpha_0\le n_0,~n_0\in \mathbb{N}$.

{\bf Definition 3.1} The function $G(t, \tau)$ that satisfies the following conditions (i) and (ii) is called {\it Green's function} for fractional differential operator $L(D_{0+})$ or IVP \eqref{9} and \eqref{10}:
\begin{align}
(i)&\quad L(D_{\tau+})G(t, \tau )=0,~t>\tau,~\tau>0,\label{11} \\
(ii)&\quad D^{\alpha_0-j}_{\tau+}G(t, \tau )|_{t=\tau}=0=\left\{
\begin{array}{rl}
1 & j=1\\
0 & j\ne 1
\end{array}\right.,
j=1, 2, \cdots,n_0,\label{12}
\end{align}
where $D^\alpha_{\tau+}$ is the Riemann-Liouville left-sided fractional differential operator with original at $t=\tau$ and $\tau$ is the parameter.

To study Green's function, now we consider IVP of FDE
\begin{align}
&L(D_{0+})y(t)=0,~t>0, \label{13}\\
&D^{\alpha_0-j}_{0+}y(0)=\left\{
\begin{array}{rl}
1 & j=1\\
0 & j\ne 1
\end{array}\right.,
j=1, 2, \cdots,n_0, \label{14}
\end{align}
and its corresponding integral equation
\begin{equation}
y(t)=\frac{t^{\alpha _0-1}}{\Gamma(\alpha_0)}-\sum_{h=1}^mI^{\alpha_0}_{0+}[a_h(t)D^{\alpha_h}_{0+}y(t)],~t>0,\label{15}
\end{equation}
where
\begin{equation}
I^{\alpha _0}_{0+}[a_h(t)D^{\alpha_h}_{0+}y(t)]=\frac{1}{\Gamma(\alpha_0)}\int _0^t\frac{a_h(\tau)D^{\alpha_h}_{0+}y(t)}{(t-\tau)^{1-\alpha_0}}d\tau ,~t>0. \label{16}
\end{equation}

{\bf Definition 3.2} For $\alpha>0$ we denote by $L^\alpha_{loc}(0, \infty)$ the set of functions $f(t)$ which fractional derivative $D^\alpha_{0+}f$ is locally integrable in the interval $(0, \infty)$, that is,
\begin{equation}
L^\alpha_{loc}(0, \infty):=\{f\in L(0, T):~D^\alpha_{0+}f\in L(0, T),~\forall T>0\}. \label{17}
\end{equation}

We need following lemma.\\

{\bf Lemma 3.1} {\it Let $y(t)\in L^{\alpha_0}_{loc}(0, \infty)$. $y(t)$ satisfies the relations \eqref{13} and \eqref{14} a. e. on $(0, \infty)$ if and only if satisfies the integral equation \eqref{15} a. e. on $(0, \infty)$}.

{\bf Proof}. First we prove the necessity. Let $y(t)\in L^{\alpha_0}_{loc}(0, \infty)$ satisfies the relations \eqref{13} and \eqref{14} a. e. on $(0, \infty)$ . We rewrite \eqref{13} in the form
\begin{equation}
(D^{\alpha_0}_{0+}y)(t)=-\sum_{h=1}^m a_h(t)(D^{\alpha_h}_{0+}y)(t), a. e. t\in (0, \infty). \label{18}
\end{equation}
Since $y(t)\in L^{\alpha_0}_{loc}(0, \infty)$, therefore $D^{\alpha_0}_{0+}y(t)\in L_{loc}(0, \infty)$, the relation \eqref{18} means that $-\sum_{h=1}^{m}a_h(t)(D^{\alpha_h}_{0+}y)(t)\in L_{loc}(0, \infty)$ a. e. on $(0, \infty)$. The relations \eqref{8} and \eqref{14} give the following
\begin{equation}
I^{\alpha_0}_{0+}D^{\alpha_0}_{0+}y(t)=y(t)-\frac{t^{\alpha_0-1}}{\Gamma(\alpha_0)}.\label{19}
\end{equation}
Applying the operator $I^{\alpha_0}_{0+}$ to both side of \eqref{18} and \eqref{14}, we obtain the equation \eqref{15}, and hence the necessity is proved.

Now we will prove the sufficiency. Let $y(t)\in L^{\alpha_0}_{loc}(0, \infty)$ satisfies \eqref{15} a. e. on $(0, \infty)$. For $j=1, 2, \cdots ,n_0$, applying the operator $D^{\alpha_0-1}_{0+}$ to both sides of \eqref{15}, we have
\begin{equation}
D^{\alpha_0-j}_{0+}y(t)=\frac{t^{j-1}}{\Gamma(j)}-\sum_{h=1}^mI^{j}_{0+}[a_h(t)D^{\alpha_h}_{0+}y(t)].\label{20}
\end{equation}
Obviously we have
\begin{equation}
\left.\frac{t^{j-1}}{\Gamma(j)}\right|_{t=0}=\left\{
\begin{array}{rl}
1 & j=1\\
0 & j\ne 1
\end{array}\right.,~~j=1,\cdots, n_0. \label{21}
\end{equation}
Since $a_h(t)D^{\alpha_h}_{0+}y(t)\in L_{loc}(0, \infty)$, we have 
\begin{equation}
I^{j}_{0+}[a_h(t)D^{\alpha_h}_{0+}y(t)]|_{t=0}=0. \label{22}
\end{equation}
Using \eqref{21}, \eqref{22} and \eqref{20}, we obtain \eqref{14}.
It is clear that
\begin{equation}
D^{\alpha_0}_{0+}\frac{t^{\alpha_0-1}}{\Gamma(\alpha_0)}=0.\label{23}
\end{equation}
Applying the operator $D^{\alpha_0}_{0+}$ to both sides of \eqref{15} and using \eqref{7} and \eqref{23}, we obtain the equation \eqref{13} and hence the sufficiency is proved.(QED)\\

 Therefore we established the equivalence of IVP of FDE \eqref{13}, \eqref{14} and integral equation \eqref{15}.

Now we find formal representation of solution of the integral equation \eqref{15} using the method of successive approximations. The successive approximations for solution to the integral equation \eqref{15} is as follows:
\begin{align}
&y_0(t)=\frac{t^{\alpha_0-1}}{\Gamma(\alpha_0)},~~y_{l+1}(t)=\frac{t^{\alpha_0-1}}{\Gamma(\alpha_0)}-\sum_{h=1}^mI^{\alpha_0}_{0+}[a_h(t)D^{\alpha_h}_{0+}y_l(t)],~l=0, 1, \cdots\\  \nonumber
&y(t)=\lim_{l\to \infty}y_l(y). \label{24}
\end{align}
Since $D^{\alpha_0}_{0+}y_0(t)=D^{\alpha_0}_{0+}\frac{t^{\alpha_0-1}}{\Gamma(\alpha_0)}=0$, it is clear that $y_0(t)\in L^{\alpha_0}_{loc}(0, \infty)$.
First approximate solution $y_1(t)$ is obtained by the following:
\begin{equation}
y_1(t)=\frac{t^{\alpha_0-1}}{\Gamma(\alpha_0)}-\sum_{h=1}^mI^{\alpha_0}_{0+}[a_h(t)D^{\alpha_h}_{0+}y_0(t)]=\frac{t^{\alpha_0-1}}{\Gamma(\alpha_0)}-\sum_{h=1}^mI^{\alpha_0}_{0+}\left[a_h(t)\frac{t^{\alpha_0-\alpha_h-1}}{\Gamma(\alpha_0-\alpha_h)}\right].\label{25}
\end{equation}
From \eqref{25}, it is clear that $y_1(t)\in L^{\alpha_0}_{loc}(0, \infty)$.
Second approximate solution $y_2(t)$ is obtained by the following :
\begin{align*}
y_2(t)&=\frac{t^{\alpha_0-1}}{\Gamma(\alpha_0)}-\sum_{h=1}^mI^{\alpha_0}_{0+}[a_h(t)D^{\alpha_h}_{0+}y_1(t)]\\
=&\frac{t^{\alpha_0-1}}{\Gamma(\alpha_0)}-\sum_{h=1}^mI^{\alpha_0}_{0+}\left\{a_h(t)D^{\alpha_h}_{0+}\left[ \frac{t^{\alpha_0-1}}{\Gamma(\alpha_0)}-\sum_{h=1}^mI^{\alpha_0}_{0+}\left(a_h(t)\frac{t^{\alpha_0-\alpha_h-1}}{\Gamma(\alpha_0-\alpha_h)}\right)\right]\right\}\\
=&\frac{t^{\alpha_0-1}}{\Gamma(\alpha_0)}-\sum_{h=1}^mI^{\alpha_0}_{0+}a_h(t) \frac{t^{\alpha_0-\alpha_h-1}}{\Gamma(\alpha_0-\alpha_h)}+\sum_{h=1}^mI^{\alpha_0}_{0+}a_h(t)D^{\alpha_h}_{0+}\sum_{h=1}^mI^{\alpha_0}_{0+}\left[a_h(t)\frac{t^{\alpha_0-\alpha_h-1}}{\Gamma(\alpha_0-\alpha_h)}\right]
\end{align*}
\begin{align}
=\frac{t^{\alpha_0-1}}{\Gamma(\alpha_0)}+\sum_{k=0}^1(-1)^{k+1}I^{\alpha_0}_{0+}\left[\sum_{h=1}^ma_h(t)I^{\alpha_0-\alpha_h}_{0+}\right]^k\sum_{h=1}^ma_h(t)\frac{t^{\alpha_0-\alpha_h-1}}{\Gamma(\alpha_0-\alpha_h)}.\label{26}
\end{align}
Here $\left[\sum_{h=1}^ma_h(t)I^{\alpha_0-\alpha_h}_{0+}\right]^k$ denotes $k$ times composition of operator $\sum_{h=1}^ma_h(t)I^{\alpha_0-\alpha_h}_{0+}$ and when $k=0$, it is unit operator.

Considering $a_h(t)\in C[0,\infty),~ \frac{t^{\alpha_0-\alpha_h-1}}{\Gamma(\alpha_0-\alpha_h)}\in L_{loc}(0,\infty)$ and \eqref{26} we have
\begin{align}
\nonumber&D^{\alpha_0}_{0+}y_2(t)=\sum_{k=0}^1(-1)^{k+1}\left[\sum_{h=1}^ma_h(t)I^{\alpha_0-\alpha_h}_{0+}\right]^k\sum_{h=1}^ma_h(t)\frac{t^{\alpha_0-\alpha_h-1}}{\Gamma(\alpha_0-\alpha_h)}\in L_{loc}(0,\infty),\\
&y_2(t)\in L^{\alpha_0}_{loc}(0,\infty).\label{27}
\end{align}
Calculating by the induction, we obtain
\begin{align}
\nonumber&y_{l+1}(t)=\frac{t^{\alpha_0-1}}{\Gamma(\alpha_0)}+\sum_{k=0}^l(-1)^{k+1}I^{\alpha_0}_{0+}\left[\sum_{h=1}^ma_h(t)I^{\alpha_0-\alpha_h}_{0+}\right]^k\sum_{h=1}^ma_h(t)\frac{t^{\alpha_0-\alpha_h-1}}{\Gamma(\alpha_0-\alpha_h)},\\
&y_{l+1}(t)\in L^{\alpha_0}_{loc}(0,\infty),~l=0,~1,\cdots .\label{28}
\end{align}
Formally taking limit as $l\rightarrow +\infty$ in the both side of \eqref{28}, the following series is obtained:
\begin{align}
y(t)=\frac{t^{\alpha_0-1}}{\Gamma(\alpha_0)}+\sum_{k=0}^\infty(-1)^{k+1}I^{\alpha_0}_{0+}\left[\sum_{h=1}^ma_h(t)I^{\alpha_0-\alpha_h}_{0+}\right]^k\sum_{h=1}^ma_h(t)\frac{t^{\alpha_0-\alpha_h-1}}{\Gamma(\alpha_0-\alpha_h)}.\label{29}
\end{align}\\

{\bf Theorem 3.1} {\it If $a_h(t)\in C[0,\infty),~h=1,\cdots,m$, then IVP of FDE \eqref{13} and \eqref{14} has a unique solution $y(t)$ in the space $L_{loc}^{\alpha_0}(0,\infty)$ and this solution is represented in the form of \eqref{29}}.

{\bf Proof}. Appling operator $D^{\alpha_0}_{0+}$ to every term of right side of the series \eqref{29}, we obtain the following series: 
\begin{align}
\sum_{k=0}^\infty(-1)^{k+1}\left[\sum_{h=1}^ma_h(t)I^{\alpha_0-\alpha_h}_{0+}\right]^k\sum_{h=1}^ma_h(t)\frac{t^{\alpha_0-\alpha_h-1}}{\Gamma(\alpha_0-\alpha_h)}.\label{30}
\end{align}

Now let us prove that this series converge in space $L(0,~T)$ for arbitrary fixed $T>0$. Let $A_h=\max_{0\leq t\leq T}|a_h(t)|,~h=1,\cdots,m~$. Using multinomial-expanding and semi-group properties of fractional integral for \eqref{30}, we can derive the following estimate: 
\begin{align}
\nonumber \sum_{k=0}^\infty&\int_0^T\left|\left[\sum_{h=1}^ma_h(t)I^{\alpha_0-\alpha_h}_{0+}\right]^k\sum_{h=1}^ma_h(t)\frac{t^{\alpha_0-\alpha_h-1}}{\Gamma(\alpha_0-\alpha_h)}\right|dt\leq\\
\nonumber &\leq\sum_{k=0}^\infty\int_0^T\left[\sum_{h=1}^m|a_h(t)|I^{\alpha_0-\alpha_h}_{0+}\right]^k\sum_{h=1}^m|a_h(t)|\frac{t^{\alpha_0-\alpha_h-1}}{\Gamma(\alpha_0-\alpha_h)}dt \\
\nonumber &\leq\sum_{k=0}^\infty\int_0^T\left[\sum_{h=1}^mA_hI^{\alpha_0-\alpha_h}_{0+}\right]^k\sum_{h=1}^mA_h\frac{t^{\alpha_0-\alpha_h-1}}{\Gamma(\alpha_0-\alpha_h)}dt= \\
\nonumber &=\sum_{k=1}^\infty\int_0^T\sum_{|\beta|=k}\frac{k~!}{\beta_1!\cdots\beta_m!}A_1^{\beta_1}\cdots A_m^{\beta_m}\frac{t^{(\alpha_0-\alpha_1)\beta_1+\cdots+(\alpha_0-\alpha_m)\beta_m-1}}{\Gamma((\alpha_0-\alpha_1)\beta_1+\cdots+(\alpha_0-\alpha_m)\beta_m)}dt\\
\nonumber &\leq\sum_{k=1}^\infty\sum_{|\beta|=k}\frac{k~!}{\beta_1!\cdots\beta_m!}A_1^{\beta_1}\cdots A_m^{\beta_m}\frac{T^{(\alpha_0-\alpha_1)\beta_1+\cdots+(\alpha_0-\alpha_m)\beta_m}}{\Gamma((\alpha_0-\alpha_1)\beta_1+\cdots+(\alpha_0-\alpha_m)\beta_m+1)}+1
\end{align}
\begin{align}
\nonumber &=\sum_{k=0}^\infty\sum_{|\beta|=k}\frac{k~!}{\beta_1!\cdots\beta_m!}A_1^{\beta_1}\cdots A_m^{\beta_m}\frac{T^{(\alpha_0-\alpha_1)\beta_1+\cdots+(\alpha_0-\alpha_m)\beta_m}}{\Gamma((\alpha_0-\alpha_1)\beta_1+\cdots+(\alpha_0-\alpha_m)\beta_m+1)}\\
&=E_{(\alpha_0-\alpha_1,\cdots,\alpha_0-\alpha_m),1}(A_1T^{\alpha_0-\alpha_1},\cdots,A_mT^{\alpha_0-\alpha_m}). \label{31}
\end{align}
Here $\beta=(\beta_1,\cdots,\beta_m)\in \mathbb{Z}_+^m$, $|\beta|=\beta_1+\cdots+\beta_m$  and $E_{(\alpha_0-\alpha_1,\cdots,\alpha_0-\alpha_m),1}(A_1T^{\alpha_0-\alpha_1}$,  $\cdots,A_mT^{\alpha_0-\alpha_m})$ is the value at $z_1=A_1T^{\alpha_0-\alpha_1}, \cdots,z_m=A_mT^{\alpha_0-\alpha_m}$ of the so - called {\it multivariate Mittag-Leffler function} $E_{(\alpha_0-\alpha_1,\cdots,\alpha_0-\alpha_m),1}(z_1,\cdots,z_m)$ (see (1.9.27) in \cite{KST}).  By the method of upper-series test, series \eqref{30} converges in the space $L(0, T)$.

Let denote sum of this series by $F(t)$, that is,
\begin{align}
F(t):=\sum_{k=0}^\infty(-1)^{k+1}\left[\sum_{h=1}^ma_h(t)I^{\alpha_0-\alpha_h}_{0+}\right]^k\sum_{h=1}^ma_h(t)\frac{t^{\alpha_0-\alpha_h-1}}{\Gamma(\alpha_0-\alpha_h)}.\label{32}
\end{align}
Then $y(t)$ of \eqref{29} can be rewritten as follows
\begin{align}
y(t)=\frac{t^{\alpha_0-1}}{\Gamma(\alpha_0)}+I^{\alpha_0}_{0+}F(t).\label{33}
\end{align}
Since $D_{0+}^{\alpha_0}y(t)=F(t)\in L(0, T)$ for any $T>0$, we have $y(t)\in L_{loc}^{\alpha_0}(0, \infty)$.

Applying the operator $D_{0+}^{\alpha_0-j}$ to both sides of \eqref{33} for $j=1,2,\cdots,n_0$, we have
\begin{align}
D_{0+}^{\alpha_0-j}y(t)=\frac{t^{j-1}}{\Gamma(j)}+I^j_{0+}F(t).\label{34}
\end{align}
Since $F(t)\in L_{loc}(0, \infty)$ , we have 
\begin{align}
I^j_{0+}F(t)|_{t=0}=0,~j=1,\cdots, n_0. \label{35}
\end{align}
By \eqref{21}, \eqref{35} and \eqref{34}, the relation \eqref{14} is obtained.

Next we will prove that $y(t)$ of \eqref{33} (or \eqref{29}) is satisfied equation \eqref{13}. From \eqref{32}, we have
\begin{align}
D_{0+}^{\alpha_0}y(t)=F(t)=\sum_{k=0}^\infty(-1)^{k+1}\left[\sum_{h=1}^ma_h(t)I^{\alpha_0-\alpha_h}_{0+}\right]^k\sum_{h=1}^ma_h(t)\frac{t^{\alpha_0-\alpha_h-1}}{\Gamma(\alpha_0-\alpha_h)}.\label{36}
\end{align}
From \eqref{33}, for $x=1,\cdots,m$, we have 
\begin{align*}
D_{0+}^{\alpha_x}&y(t)=\frac{t^{\alpha_0-\alpha_x-1}}{\Gamma(\alpha_0-\alpha_x)}+I^{\alpha_0-\alpha_x}_{0+}F(t)=\\
=&\frac{t^{\alpha_0-\alpha_x-1}}{\Gamma(\alpha_0-\alpha_x)}+\sum_{k=0}^\infty(-1)^{k+1}I^{\alpha_0-\alpha_x}_{0+}\left[\sum_{h=1}^ma_h(t)I^{\alpha_0-\alpha_h}_{0+}\right]^k\sum_{h=1}^ma_h(t)\frac{t^{\alpha_0-\alpha_h-1}}{\Gamma(\alpha_0-\alpha_h)}
\end{align*}
and hence we have
\begin{align*}
\nonumber \sum&_{x=1}^{m}a_x(t)D_{0+}^{\alpha_x}y(t)=\\
\nonumber =&\sum_{x=1}^{m}a_x(t)\frac{t^{\alpha_0-\alpha_x-1}}{\Gamma(\alpha_0-\alpha_x)}+\sum_{k=0}^\infty(-1)^{k+1}\left[\sum_{h=1}^ma_h(t)I^{\alpha_0-\alpha_h}_{0+}\right]^{k+1}\sum_{h=1}^ma_h(t)\frac{t^{\alpha_0-\alpha_h-1}}{\Gamma(\alpha_0-\alpha_h)}\\
\nonumber =&\sum_{x=1}^{m}a_x(t)\frac{t^{\alpha_0-\alpha_x-1}}{\Gamma(\alpha_0-\alpha_x)}+\sum_{k=1}^\infty(-1)^{k}\left[\sum_{h=1}^ma_h(t)I^{\alpha_0-\alpha_h}_{0+}\right]^{k}\sum_{h=1}^ma_h(t)\frac{t^{\alpha_0-\alpha_h-1}}{\Gamma(\alpha_0-\alpha_h)}
\end{align*}
\begin{align}
=\sum_{k=0}^\infty(-1)^{k}\left[\sum_{h=1}^ma_h(t)I^{\alpha_0-\alpha_h}_{0+}\right]^{k}\sum_{h=1}^ma_h(t)\frac{t^{\alpha_0-\alpha_h-1}}{\Gamma(\alpha_0-\alpha_h)}. \label{37}
\end{align}
From \eqref{36} and \eqref{37}, we obtain $=D^{\alpha_0}_{0+}y(t)+\sum_{x=1}^{m}a_x(t)D_{0+}^{\alpha_x}y(t)=0$. Thus $y(t)$ of  of \eqref{33} (or \eqref{29}) satisfies the equation \eqref{13}.

By corollary 3.6 of \cite{KST}, we obtain the uniqueness result for the IVP \eqref{13} and \eqref{14}. This completes the proof of Theorem 3.1. (QED)\\

{\bf Corollary 3.1} {\it Let $a_h(t)=A_h=const, h=1,\cdots,m$. Then the unique solution $y(t)\in L_{loc}^{\alpha_0}(0,\infty)$ of the IVP \eqref{13} and \eqref{14} is represented by}
\begin{align}
y(t)=\sum_{k=0}^\infty(-1)^{k}\sum_{|\beta|=k}\frac{k~!~A_1^{\beta_1}\cdots A_m^{\beta_m}~\cdot~t^{(\alpha_0-\alpha_1)\beta_1+\cdots+(\alpha_0-\alpha_m)\beta_m+\alpha_0-1}}{\beta_1!\cdots\beta_m!~\Gamma[(\alpha_0-\alpha_1)\beta_1+\cdots+(\alpha_0-\alpha_m)\beta_m+\alpha_0]}.\label{38}
\end{align}

{\bf Proof}. Let  $a_h(t)=A_h=const, h=1,\cdots,m$ in the solution representation \eqref{29} of IVP \eqref{13} and \eqref{14} and use the semi-group properties of fractional integral and multi-term's expanding. Then the discussion similar with the derivation of \eqref{31} gives \eqref{38}.(QED)\\

{\bf Remark 3.1} The representation $y(t)$ of \eqref{38} is coincided with multivariate {\it Mittag-Leffler} function $E_{(\alpha_0-\alpha_1,\cdots,\alpha_0-\alpha_m),\alpha_0}(-A_1t^{\alpha_0-\alpha_1}$,$\cdots,-A_mt^{\alpha_0-\alpha_m})$   (See (1.9.27) of \cite{KST}). Note that multivariate {\it Mittag-Leffler} function was introduced originally by Y. Luchko.\\

{\bf Remark 3.2} Although the solutions \eqref{29} and \eqref{38} of IVP \eqref{13} and \eqref{14} are series expression but give an algorithm for calculation of the solution directly.\\

{\bf Corollary 3.2} {\it Let $a_h(t)=A_h=const, h=1,\cdots,m$. Then the unique solution $y(t)\in L_{loc}^{\alpha_0}(0,\infty)$ of the IVP \eqref{13} and \eqref{14} is represented by Mittag-Leffler function of two parameters as follows}:
\begin{align}
\nonumber y(t)=\sum_{l=0}^\infty\frac{(-1)^{l}}{l!}\sum_{\beta_2+\cdots+\beta_m=l}\frac{l!~\prod_{i=2}^mA_i^{\beta_i}}{\beta_2!\cdots\beta_m!}\cdot t^{(\alpha_0-\alpha_1)l+\alpha_0+\sum_{j=2}^m(\alpha_1-\alpha_j)\beta_j-1}\cdot\\
\cdot E_{\alpha_0-\alpha_1,\alpha_0+\sum_{j=2}^m(\alpha_1-\alpha_j)\beta_j-1}^{(l)}(-A_1t^{\alpha_0-\alpha_1}).\label{39}
\end{align}
Here
\[E_{\alpha, \beta}^{(l)}(z):=\sum_{i=0}^{\infty}\frac{(i+l)!}{i!}\frac{z^{i}}{\Gamma(\alpha i+\alpha l+\beta)}.\]

{\bf Proof}. Let $\beta_2+\cdots+\beta_m=l$ for multi-index $\beta=(\beta_1,\beta_2,\cdots,\beta_m)$ in \eqref{38}. Then $k=|\beta|=\beta_1+l$, and therefore we rewrite \eqref{38} as the form $$\sum_{k=0}^\infty(-1)^{k}\sum_{|\beta|=k}\frac{k~!~}{\beta_1!\cdots\beta_m!}\cdots=\sum_{l=0}^\infty\sum_{\beta_1=0}^\infty\frac{(-1)^{l+\beta_1}}{l!}\sum_{\beta_2+\cdots+\beta_m=l}\frac{l!~(\beta_1+l)!}{\beta_2!\cdots\beta_m!\beta_1!}\cdots$$ 
and consider $$(\alpha_0-\alpha_1)\beta_1+\cdots+(\alpha_0-\alpha_m)\beta_m+\alpha_0=(\alpha_0-\alpha_1)\beta_1+(\alpha_0-\alpha_1)l+\alpha_0+\sum_{j=2}^m(\alpha_1-\alpha_j)\beta_j,$$ 
then we can easily obtain \eqref{39}.(QED)\\

{\bf Remark 3.3} In \cite{MS1, pod1, pod2} $y(t)$ represented by \eqref{38} or \eqref{39} is called Green's function of IVP \eqref{9} and \eqref{10} in the case with constant coefficients.\\

{\bf Theorem 3.2}. {\it If $a_h(t)\in C[0,\infty),~h=1,\cdots,m$, then there exists unique Green's function $G(t,\tau)$ of the fractional differential operator $L(D_{0+})$ (solution of IVP \eqref{11} and \eqref{12} in the space $L_{loc}^{\alpha_0}(\tau,\infty)$ and it is represented as follows:}
\begin{align}
\nonumber G(t,&\tau)=\frac{(t-\tau)^{\alpha_0-1}}{\Gamma(\alpha_0)}+\\
+&\sum_{k=0}^\infty(-1)^{k+1}I^{\alpha_0}_{\tau+}\left[\sum_{h=1}^ma_h(t)I^{\alpha_0-\alpha_h}_{\tau+}\right]^k\sum_{h=1}^ma_h(t)\frac{(t-\tau)^{\alpha_0-\alpha_h-1}}{\Gamma(\alpha_0-\alpha_h)},~t>\tau>0.\label{40}
\end{align}
{\it In particular, if $a_h(t)=A_h=const, h=1,\cdots,m$, then we have}
\begin{align*}
G(t,\tau)=\sum_{k=0}^\infty(-1)^{k}\sum_{|\beta|=k}\frac{k~!~A_1^{\beta_1}\cdots A_m^{\beta_m}~\cdot~(t-\tau)^{(\alpha_0-\alpha_1)\beta_1+\cdots+(\alpha_0-\alpha_m)\beta_m+\alpha_0-1}}{\beta_1!\cdots\beta_m!~\Gamma[(\alpha_0-\alpha_1)\beta_1+\cdots+(\alpha_0-\alpha_m)\beta_m+\alpha_0]}.
\end{align*}

{\bf Proof}. The proof of theorem 3.2 is similar to that of theorem 3.1.\\

Using Green's functions, we can obtain representation of solutions to inhomogeneous IVP. The following theorem holds.\\

{\bf Theorem 3.3} {\it If $a_h(t)\in C[0,\infty),~h=1,\cdots,m$, then there exists unique solution $y(t)\in I_{0+}^{\alpha_0}(L)$ (when $a=0, b=T, \forall T>0$; see definition 2.5) of the IVP of FDE \eqref{9} and \eqref{10} and the solution is represented in the following form}:
\begin{align}
\nonumber y&(t)=\int_0^tG(t,\tau)h(\tau)d\tau=\\
\nonumber &=\int_0^t\left\{\frac{(t-\tau)^{\alpha_0-1}}{\Gamma(\alpha_0)} \right.+\\
&\quad+\left.\sum_{k=0}^\infty(-1)^{k+1}I^{\alpha_0}_{\tau+}\left[\sum_{h=1}^ma_h(t)I^{\alpha_0-\alpha_h}_{\tau+}\right]^k\sum_{h=1}^ma_h(t)\frac{(t-\tau)^{\alpha_0-\alpha_h-1}}{\Gamma(\alpha_0-\alpha_h)}\right\}h(\tau)d\tau.  \label{41}
\end{align}
{\it Here $G(t,\tau)$ was given in \eqref{40}}. 

{\bf Proof}. Using the definition of fractional integral and Fubini's theorem, we can prove the equality \eqref{41}. Then similarly with theorem 3.1, if we substitute \eqref{41} to the IVP \eqref{9} and \eqref{10}, then we can prove that $y(t)$ is the solution of \eqref{9} and \eqref{10}. (QED)
\section{Examples}
First let consider the Green's function of the fractional differential operator 
\begin{align}
L(D_{0+})=D_{0+}^{1.5}+tD_{0+}^{0.5}. \label{42}
\end{align}
In this case $\alpha_0=1.5,~\alpha_1=0.5, n_0=2,~a_1(t)=t$. By \eqref{40}, its Green function is as follows:
\begin{align}
G(t,\tau)=\frac{(t-\tau)^{0.5}}{\Gamma(1.5)}+\sum_{k=0}^\infty(-1)^{k+1}I^{1.5}_{\tau+}\left[tI^{1}_{\tau+}\right]^kt~,~~t>\tau>0~.\label{43}
\end{align}
If we substitute \eqref{42} and \eqref{43} into \eqref{11} and \eqref{12}, then we can know that the $G(t,\tau)$ is the Green function of \eqref{42}. Now we calculate some terms of \eqref{43}. In the term $I^{1.5}_{\tau+}t=\frac{1}{\Gamma(1.5)}\int_{\tau}^t(t-\xi)^{0.5}\xi d\xi$ of the series when $k=0$, using change of variable $t=\xi+s(t-\tau)$, the interval of integral $[\tau, t]$ is changed into $[0, 1]$ and by the simple calculation we have 
\[I^{1.5}_{\tau+}t=\frac{1}{1.5\Gamma(1.5)}t(t-\tau)^{1.5}-\frac{1}{2.5\Gamma(2.5)}t(t-\tau)^{2.5}.\]
In the second term, $tI^{1}_{\tau+}t=t\int_{\tau}^t\xi d\xi=t^3/2-t\tau^2/2$ and thus we have \[I^{1.5}_{\tau+}tI^{1}_{\tau+}t=\frac{1}{\Gamma(1.5)}\int_{\tau}^t(t-\xi)^{0.5}(\xi^3/2-\xi\tau^2/2)d\xi.\] 
This integral is easily calculated using the similar method as the above. Thus we have the series representation of Green's function:
\begin{align*}
G(t,\tau)=\frac{(t-\tau)^{0.5}}{\Gamma(1.5)}+\frac{t(t-\tau)^{1.5}}{1.5\Gamma(1.5)}-\frac{t(t-\tau)^{2.5}}{2.5\Gamma(2.5)}+\cdots~,~~t>\tau>0~.
\end{align*}

Now using the formula \eqref{41}, let solve the following IVP:
\begin{align}
&D_{0+}^{1.5}y(t)+t^2D_{0+}^{0.5}y(t)+t^3y(t)=\frac{t^{-0.8}}{\Gamma(0.2)}, \label{44}\\
&(D_{0+}^{1.5-k}y)(0+)=0,~k=1,2.  \label{45}
\end{align}
By theorem 3.3, this problem has unique solution $y\in I^{1.5}(L)$. In this case
$\alpha_0=1.5,~\alpha_1=0.5,~\alpha_2=0,~n_0=2,~m=2,~a_1(t)=t^2,~a_2(t)=t^3$ and $h(t)=t^{-0.8}/{\Gamma(0.2)}\in L_{loc}(0,\infty)$. Therefore using \eqref{41}, we have
\begin{align*}
y(t)=\int_0^t&\left\{\frac{(t-\tau)^{0.5}}{\Gamma(1.5)} \right.+\\
+&\left.\sum_{k=0}^\infty(-1)^{k+1}I^{1.5}_{\tau+}\left[t^2I^{1}_{\tau+}+t^3I^{1.5}_{\tau+}\right]^k\left(t^2\frac{(t-\tau)^{0}}{\Gamma(1)}+t^3\frac{(t-\tau)^{0.5}}{\Gamma(1.5)}\right)\right\}h(\tau)d\tau.  \label{41}
\end{align*}
If we calculate it using the similar way as the above, we have the following series representation of the IVP \eqref{44} and \eqref{45}: 
\begin{align*}
y(t)=&\frac{t^{0.7}}{\Gamma(1.7)}-\left[\frac{\Gamma(3.2)}{\Gamma(1.2)}\frac{t^{3.7}}{\Gamma(4.7)}+\frac{\Gamma(4.7)}{\Gamma(1.7)}\frac{t^{5.2}}{\Gamma(6.2)}\right]+\left[\frac{\Gamma(3.2)}{\Gamma(1.2)}\frac{\Gamma(6.2)}{\Gamma(4.2)}\frac{t^{6.7}}{\Gamma(7.7)}\right.\\
+&\left.\left(\frac{\Gamma(3.2)}{\Gamma(1.2)}\frac{\Gamma(7.7)}{\Gamma(4.7)}+\frac{\Gamma(4.7)}{\Gamma(1.7)}\frac{\Gamma(7.7)}{\Gamma(5.7)}\right)\frac{t^{8.2}}{\Gamma(9.2)}+\frac{\Gamma(4.7)}{\Gamma(1.7)}\frac{\Gamma(9.2)}{\Gamma(6.2)}\frac{t^{9.7}}{\Gamma(10.7)}\right]-\cdots.
\end{align*}

\section{Conclusions}
In this paper we presented an explicit representation formula for the Green's function of the general linear fractional differential operator with continuous variable coefficients, in the meaning of Riemann-Liouville and showed that this result is consistent with previous results in the case with constant coefficients. The representation formula of the Green’s function for linear fractional differential operator with continuous variable coefficients will be used as a powerful tool to solve the Caputo fractional differential equations as well as Riemann-Liouville fractional equations.\\

{\bf Acknowledgement}: Authors would like to thank the editor-in-chief A. M. A. El-Sayed and anonymous reviewers' help and advice.


\begin{thebibliography}{99}

\bibitem{BRT} Bonilla B., Rivero M. and Trujillo J.J., \textit{On systems linear fractional differential equations with constant coefficients}, Applied Mathematics and Computation,187, 68-78, 2007, \href{http://dx.doi.org/10.1016/j.amc.2006.08.104}{DOI: 10.1016/j.amc.2006.08.104}

\bibitem{HLT} Hilfer R.,Luchko Y. and Tomovski Z., {\it Operational method for the solution of fractional differential equations with generalized Riemann-Liouville fractional derivatives}, Fract. Calc. Appl. Anal., 12, 3, 299-318, 2009, \href{http://www1.beuth-hochschule.de/~luchko/papers/Hilfer_Luchko_Tomovski_FCAA_12_3_printed_version.pdf}{cross-ref}

\bibitem{HLL} Hu Y., Luo Y., Lu Z., {\it Analytical Solution of the linear fractional differential equation by Adomian decomposition method}, J. Comput. Appl. Math., Vol. 215, Issue 1, 15 May 2008, 220-229, \href{http://dx.doi.org/10.1016/j.cam.2007.04.005}{cross-ref}

\bibitem{KRRT} Kilbas A.A.,Rivero M.,Rodrignez-Germa L.,Trujillo J.J., $\alpha$-{\it Analytic solutions o some linear fractional differential equations with variable coefficients}, Appl. Math. and Comput.,187, 239-249, 2007, \href{http://dx.doi.org/10.1016/j.amc.2006.08.121}{cross-ref}

\bibitem{KST} Kilbas A.A., Srivastava H.M. and Trujillo, J. J., Theory and Applications of Fractional Differential Equations, Elsevier, Amsterdam-Tokyo, 2006

\bibitem{mes} Meshkov S.I., Viscoelastic Properties of metals, Metallurgia, Moscow, 1974

\bibitem{MR} Miller K.S. and Ross B., An Introduction to the Fractional Calculus and Fractional Differential Equations, Wiley and Sons, New York, 1993

\bibitem{MS1} Morita T. and Sato K., {\it Neumann-Series Solution of Fractional Differential Equation}, Interdisciplinary Information Sciences, Vol.16, No.1, 2010, 127-137, \href{http://dx.doi.org/10.4036/iis.2010.127}{DOI: 10.4036/iis.2010.127}

\bibitem{MS2} Morita T. and Sato K., {\it Solution of Fractional Differential Equation in Terms of Distribution Theory}, Interdisciplinary Information Sciences, Vol.12, No.2, 2006, 71-83, \href{http://dx.doi.org/10.4036/iis.2006.71}{DOI: 10.4036/iis.2006.71} 

\bibitem{nai} Naimark M. A., Linear Differential Operators, Nayka, Moscow, 1969

\bibitem{pod1} Podlubny I., Fractional Differential Equations, Academic Press, San Diego, 1999

\bibitem{pod2} Podlubny I., {\it The Laplace Transform Method for Linear Differential Equations of the Fractional order}, Inst. Exp. Phys., Slovak Acad. Sci. No UEF-02-94, 1994, Kosice, 1-32, \href{http://arxiv.org/pdf/funct-an/9710005.pdf}{cross-ref}

\bibitem{HL} Huang X., Lu X., {\it The use of fractional B-splines wavelets in Multi-terms fractional ordinary Differential Equations}, International Journal of Differential Equations, Volume 2010 (2010), Article ID 968186, 13p, \href{http://dx.doi.org/10.1155/2010/968186}{DOI:10.1155/2010/968186}

\bibitem{SKM} Samko S.G., Kilbas A.A., Marichev O.I., Fractional Integrals and Derivatives: Theory and Applications, New York and London, Gordon and Breach Science Publishers. 1993.

\end{thebibliography}
\end{document}